\documentclass[a4paper,11pt]{article}

\usepackage{contribution}
\newcommand{\weblink}[2][]{%
    \ifthenelse{\equal{#1}{}}%
    {\textnormal{\url{#2}}}%
    {\textnormal{\href{#2}{#1}}}%
}

\newcommand{\acknowledgements}[1]{%
  \bigskip\bigskip
  \textsf{\textbf{\Large Acknowledgements}} \\[2ex]
  {#1}
  \bigskip
}

\def\beq{\begin{equation}}
\def\eeq#1{\label{#1}\end{equation}}
\def\eeqn{\end{equation}}

\def\beqa{\begin{eqnarray}}
\def\eeqa#1{\label{#1}\end{eqnarray}}
\def\eeqan{\end{eqnarray}}

\let\bar=\overbar

\def\Dslash{\not{\hbox{\kern-4pt $D$}}}
\def\dslash{\not{\hbox{\kern-2pt $\del$}}}

\def\msb{{\bar{\ssstyle M \kern -1pt S}}}

\newcommand{\contribution}[7][]{%
  \clearpage
  \thispagestyle{plain}
  \ifthenelse{\equal{#1}{}}
  {\hypersetup{pdftitle={#2}}}
  {\hypersetup{pdftitle={#1}}}
  \hypersetup{pdfauthor={{#3} {#4}}}
  {\centering\normalfont\LARGE\bfseries\sffamily #2 \par\nobreak}
  \lhead{}
  \chead{%
    \textit{\footnotesize XIV International Conference on Hadron Spectroscopy
      (\weblink[\textit{hadron2011}]{http://www.hadron2011.de}), 13-17 June 2011, Munich, Germany}%
  }
  \rhead{}
  \bigskip
  \begin{center}
    {#3} {#4}\ifthenelse{\equal{#6}{}}{}{\footnote{\weblink[#6]{mailto:#6}}}
    \ifthenelse{\equal{#7}{}}{}{#7} \\
    \textit{#5}
  \end{center}
  \bigskip
}

\renewcommand{\abstract}[1]{%
  \begin{center}
    \begin{minipage}{0.85\textwidth}
      \begin{footnotesize}
        #1
      \end{footnotesize}
    \end{minipage}
  \end{center}
  \bigskip
}

\begin{document}

% % % % % % % % % % % % % % % % % % % % % % % % % % % % % % % % % % % % % % % % %
% your proceedings
%%%%%%%%%%%%%%%%%%%%%%%%%%%%%%%%%%%%%%%%%%%%%%%%%%%%%%%%%%%%%%%%%%%%%%%%%%%%%%%%%
%
% template for hadron2011 contribution
%
% please do not rename this file
%
% to create document run
%
%     pdflatex hadron2011.tex
%
%%%%%%%%%%%%%%%%%%%%%%%%%%%%%%%%%%%%%%%%%%%%%%%%%%%%%%%%%%%%%%%%%%%%%%%%%%%%%%%%%
{  % do not remove

%%%%%%%%%%%%%%%%%%%%%%%%%%%%%%%%%%%%%%%%%%%%%%%%%%%%%%%%%%%%%%%%%%%%%%%%%%%%%%%%%
% please define your macros here
\makeatletter
\@ifundefined{c@affiliation}%
{\newcounter{affiliation}}{}%
\makeatother
\newcommand{\affiliation}[2][]{\setcounter{affiliation}{#2}%
\ensuremath{{^{\alph{affiliation}}}\text{#1}}}

%
%%%%%%%%%%%%%%%%%%%%%%%%%%%%%%%%%%%%%%%%%%%%%%%%%%%%%%%%%%%%%%%%%%%%%%%%%%%%%%%%%

%%%%%%%%%%%%%%%%%%%%%%%%%%%%%%%%%%%%%%%%%%%%%%%%%%%%%%%%%%%%%%%%%%%%%%%%%%%%%%%%%
% define title, author, and address
% contribution[short title]{title}{author first name}{author last name}{author address}{author email}{collaboration}
% the short title will appear in the page headers and the TOC of the book of proceedings
% the last two arguments may be left empty
\contribution[Molecular effects in Charmonium Spectrum]  % short title (optional)
{Molecular effects in Charmonium Spectrum}  % title
{David R.}{Entem}  % first and last name of author
{\affiliation[Departamento de F\'\i sica Fundamental and IUFFyM]{1} \\
Universidad de Salamanca, E-37008 Salamanca, Spain}  % author address
{entem@usal.es}  % author email optional
{\!\!$^,\affiliation{1}$, 
Pablo G. Ortega\affiliation{1}, 
and Francisco Fern\'andez\affiliation{1}}  % collaboration (optional)
%
%%%%%%%%%%%%%%%%%%%%%%%%%%%%%%%%%%%%%%%%%%%%%%%%%%%%%%%%%%%%%%%%%%%%%%%%%%%%%%%%%

%%:wq
%%%%%%%%%%%%%%%%%%%%%%%%%%%%%%%%%%%%%%%%%%%%%%%%%%%%%%%%%%%%%%%%%%%%%%%%%%%%%%%
% abstract
\abstract{%
We study the influence of possible molecular structures in the
charmonium spectrum. We focus on the $0^{++}$ and $1^{--}$ sectors.
In the first one we coupled the $2 ^3P_0$ $q\bar q$ pair with
$DD$, $J/\psi \omega$, $D_sD_s$ and $J/\psi\phi$ channels and
we obtain two states compatibles with the $X(3945)$ and the $Y(3940)$.
In the second one we include the $3^3S_1$ and $2^3D_1$ charmonium states
coupled to $DD$, $DD^*$, $D^*D^*$, $D_sD_s$, $D_sD_s^*$ and $D_s^*D_s^*$.
In this calculation we obtain a new molecular state that could be the $G(3900)$ or
the controversial $Y(4008)$ and two $c\bar c$ states dressed by the
molecular components assigned to the $\psi(4040)$ and
the $\psi(4160)$. The two $c\bar c$ states show interesting properties and
in particular they solve the strong disagreement of the decay branching
ratios measured by BABAR.
}
%
%%%%%%%%%%%%%%%%%%%%%%%%%%%%%%%%%%%%%%%%%%%%%%%%%%%%%%%%%%%%%%%%%%%%%%%%%%%%%%%%%

%%%%%%%%%%%%%%%%%%%%%%%%%%%%%%%%%%%%%%%%%%%%%%%%%%%%%%%%%%%%%%%%%%%%%%%%%%%%%%%%%
% main text
% for short contributions sections are optional
\section{Introduction}

The discovery of the $J/\psi$ meson in 1974 was the experimental confirmation of
the existence of the charm quark introduced theoretically in 1970 by Glashow, 
Iliopoulus and Maiani to explain the cancellation of loop diagrams in $K^0$ weak decays.
Consisting of a charm $c$ quark and a $\bar c$ antiquark, the $J/\psi$ particle became
the starting point of a whole family of bound states called charmonium. A further
milestone in the knowledge of the charmonium structure began in 2002 with the new
data coming from high luminosity experiments at $B$ factories. Since then,
many new states have been observed. A summary of these states can be found
in Ref.~\cite{Brambilla}. Most of them are difficult to understand
in a quark-antiquark framework and meson-antimeson molecular states may represent
an alternative explanation to these states.

Probably the most popular of these new states is the $X(3872)$ which lies
very close to the $DD^*$ threshold and is the most accepted candidate to be
a meson-antimeson bound state. In Ref.~\cite{Ortega10} we have performed a
calculation of the $X(3872)$ state as a $DD^*$ molecule in the framework
of a constituent quark model~\cite{Vijande}. There also the coupling to
$q\bar q$ states is included and only when we mix the $DD^*$ channel and
the $\chi_{c1}(2P)$ state the $X(3872)$ appears as a bound state. The
original $\chi_{c1}(2P)$ $q\bar q$ state acquires a significant $DD^*$
component and can be identified with the $X(3940)$.

Following these ideas we have started a program to study the influence
of possible molecular structures in the charmonium spectrum. We have
generalized our formalism to study resonance above thresholds. In
this contribution we focus on the results found in the $0^{++}$ and
$1^{--}$ sectors.

\section{Theoretical framework}

We work in the framework of a coupled channel calculation
following Ref.~\cite{Baru} generalized to include several
meson-antimeson channels and several $q\bar q$ states.
The $T$-matrix is decomposed in a non-resonant and a
resonant contribution.
We look for poles of the $T$-matrix in the second
Riemann sheet looking for zeros of the dressed
propagator. Then we solve an eigenvalue problem to
obtained the $c\bar c$ amplitudes and with them
and the dressed vertex we obtained the meson-antimeson
wave functions. Finally we define the partial widths
evaluating the residues of the pole.

The $q\bar q$ states are found solving the Schr\"odinger
equation with the $q\bar q$ interaction of Ref.~\cite{Vijande}.
The meson-antimeson interaction is consistently obtained
using the Resonating Group Method with these wave functions
and interaction.
Finally we couple the $q\bar q$ states with the $q\bar q q\bar q$ states
using the microscopic $^3 P_0$ model. 

%%%%%%%%%%%%%%%%%%%%%%%%%%%%%%%%%%%%%%%%%%%%%%%%%%%%%%%%%%%%%%%%%%%%%%%%%%%%%%%%%

\section{The $0^{++}$ sector}

We study the $3900$ MeV energy region and we include in the calculation the
$\chi_{c0}(2P)$ $c\bar c$ state and the $DD$, $J/\psi\omega$, $D_sD_s$ and
$J/\psi\phi$ meson-antimeson states. We found two states corresponding to
the first two lines in Table~\ref{tab:0++}. For comparison we show the 
result of the updated Cornell model ($C^3$) of Ref.~\cite{Eichten05}.
Our second state corresponds to the $\chi_{c0}(2P)$ dressed by the
meson-antimeson states and is similar to the one found in the $C^3$
model. In addition our framework also allows to study resonances due 
to the binding of meson-antimeson channels and we found and additional
state. 

The experimental situation is not clear. Two states were found in these
energy region, namely the $X(3945)$~\cite{Uehara} and the $Y(3940)$~\cite{Choi}
and they could be the two states we find, although recently these two states
have been summarize in one, the $X(3915)$, due to compatible properties.

\begin{table}[tb]
\begin{center}
\begin{tabular}{c|ccccc|ccc}
Mass(MeV) & $2^3P_0$ & $DD$ & $J/\psi\omega$ & $D_sD_s$ & $J/\psi\phi$ 
& $\Gamma_{DD}$ & $\Gamma_{J/\psi\omega}$ & $\Gamma_{D_sD_s}$ \\
\hline
$3896.05 -i 2.10$ & $34.22$ & $46.67$ & $9.41$ & $9.67$ & $0.03$ & $3.37$ & $0.83$ & $-$ \\
$3970.07 -i 94.67$ & $57.27$ & $35.32$ & $0.15$ & $5.72$ & $1.54$ & $38.69$ & $2.89$ & $147.76$ \\
\hline
$3881.4-i\,30.75$ & $49$ & $34.22$ & $-$ & $4.41$ & $-$ \\
\end{tabular}
\caption{\label{tab:0++} Pole position, probabilities and partial widths for the
two states (in the first two lines) obtained in the $0^{++}$ sector. 
For comparison we show in the last line the result of
the updated Cornell model ($C^3$) of Ref.~\cite{Eichten05} for the
only state found in the $3900$ MeV energy region.}
\end{center}
\end{table}

\section{The $1^{--}$ sector}

We study the $4100$ MeV energy region and we include the $3 ^3S_1$ and
$2 ^3D_1$ $c\bar c$ states and the $DD$, $DD^*$, $D^*D^*$, $D_sD_s$, $D_sD_s^*$
and $D_s^*D_s^*$ meson-antimeson channels. Our results are summarized in
Tables~\ref{tab:1--} and \ref{tab:gam1--}. We find two states corresponding
to the dressing of the original $c\bar c$ states corresponding to the
well established $\psi(4040)$ and $\psi(4160)$ and one additional state
that could be the $G(3900)$ or the controversial $Y(4008)$.

\begin{table}[tb]
\begin{center}
\begin{tabular}{c|cccccccc}
$M\,(MeV)$ & $3 ^3S_1$ & $2 ^3D_1$ &
$DD$ & $DD^*$ & $D^*D^*$
& $D_sD_s$ & $D_sD_s^*$ &
$D_s^*D_s^*$ \\
\hline
$3994.6-i\,11.60$ & $31.56$ & $3.00$ & $2.49$ & $36.44$ & $17.75$ &
$7.53$ & $0.52$ & $0.71$ \\
$4048.4-i\,7.54$  & $0.92$  & $36.15$& $2.99$ & $23.49$ & $25.81$ &
$8.86$ & $0.92$ & $0.85$ \\
$4123.9-i\,71.11$ & $59.01$ & $0.98$ & $2.13$ & $6.84$  & $19.19$ &
$0.75$ & $3.37$ & $7.73$ \\
\hline
$4038-i \, 37$ & $44.89$ & $0.16$ & $2.87$ & $20.36$ & $23.10$ & $0.98$ & $1.58$ & $1.08$ \\
$(4160)-i\,24.6$ & $0.09$ & $47.61$ & $8.37$ & $4.24$ & $8.87$ & $0.55$ & $0.96$ & $1.31$ \\
\end{tabular}
\caption{\label{tab:1--} Pole position and probabilities for the
three states (in the first three lines) obtained in the $1^{--}$ sector. 
For comparison we show in the last line the result of
the updated Cornell model ($C^3$) of Ref.~\cite{Eichten05} for the
two states found in the $4100$ MeV energy region.}
\end{center}
\end{table}

\begin{table}[tb]
\begin{center}
\begin{tabular}{cc|ccccc}
$M$  &$\Gamma$ & $\Gamma(DD)$ & $\Gamma(DD^*)$ & $\Gamma(D^*D^*)$ &
$\Gamma(D_sD_s)$& $\Gamma(D_sD_s^*)$  \\
\hline
$3994.6$ & $23.37$  & $0.12$ & $19.09$ & $-$     & $4.16$  & $-$     \\
$4048.4$ & $15.09$  & $0.51$ & $7.24$  & $4.42$  & $2.92$  & $-$     \\
$4123.9$ & $142.23$ & $4.73$ & $7.51$  & $100.03$& $3.82$  &$26.15$  \\
\end{tabular}
\caption{\label{tab:gam1--} Partial widths for the
three states obtained in the $1^{--}$ sector.}
\end{center}
\end{table}

The dressed states shows a peculiar feature. Although originally the $3^3S_1$
state is below the $2^3D_1$, when coupled to the meson-antimeson channels, the
lowest state is predominantly $2^3D_1$ and the highest $3^3S_1$. This is due
to the additional state which has a stronger coupling with the $3^3S_1$ and push
the dress state to higher energies. In order to test this new assignment we
have calculated the branching ratios measured by BABAR~\cite{Aubert}. We
show the results for the original bare $q\bar q$ states and the states
obtained in the coupled channel calculation in Table~\ref{tab:branch}.
There is a strong disagreement with the experimental data for the bare states,
while the results for the coupled channel calculation are within three standard
deviations.

\begin{table}[tb]
\begin{center}
\begin{tabular}{cccc}
Ratio  & Experimental value & $q\bar q$ with $^3P_0$ & Coupled channel
\\
\hline
$\frac{\mathcal{B}(\psi(4040)\to D\bar D)}{\mathcal{B}(\psi(4040)\to
D\bar D^*)}$     
    & $0.24\pm 0.05\pm 0.12$ & $0.21$ & $0.07$ \\ 
$\frac{\mathcal{B}(\psi(4040)\to D^*\bar
D^*)}{\mathcal{B}(\psi(4040)\to D\bar D^*)}$ 
    & $0.18\pm 0.14\pm 0.03$ & $3.7$ & $0.61$ \\ 
\hline
    $\frac{\mathcal{B}(\psi(4160)\to D\bar D)}{\mathcal{B}(\psi(4160)\to
D^*\bar D^*)}$   
   & $0.02\pm 0.03\pm 0.02$ & $0.27$ & $0.05$ \\ 
   $\frac{\mathcal{B}(\psi(4160)\to D\bar
   D^*)}{\mathcal{B}(\psi(4160)\to D^*\bar D^*)}$ 
    & $0.34\pm 0.14\pm 0.05$ & $0.027$ & $0.08$ \\ 
\end{tabular}
\caption{\label{tab:branch} Branching ratios measured by BABAR~\cite{Aubert}
for the $\psi(4040)$ and the $\psi(4160)$ (second column). The third column
shows the result for the bare $c\bar c$ states and the fourth the results
of the states of Table~\ref{tab:1--}.}
\end{center}
\end{table}

%%%%%%%%%%%%%%%%%%%%%%%%%%%%%%%%%%%%%%%%%%%%%%%%%%%%%%%%%%%%%%%%%%%%%%%%%%%%%%%%%
% acknowledgements (optional)
\acknowledgements{%
This work has been partially funded by Ministerio de Ciencia y Tecnolog\'ia under Contract No. FPA2010-21750-C02-02, by the European Community-Research Infrastructure Integrating Activity 'Study of Strongly Interacting Matter' (HadronPhysics2 Grant No. 227431) and by the Spanish Ingenio-Consolider 2010 Program CPAN (CSD2007-00042).
}

%%%%%%%%%%%%%%%%%%%%%%%%%%%%%%%%%%%%%%%%%%%%%%%%%%%%%%%%%%%%%%%%%%%%%%%%%%%%%%%%%
% bibliographic items can be constructed using the LaTeX format in SPIRES
% see http://www.slac.stanford.edu/spires/hep/latex.html
% SPIRES will also supply the CITATION line information; please include it

%
%%%%%%%%%%%%%%%%%%%%%%%%%%%%%%%%%%%%%%%%%%%%%%%%%%%%%%%%%%%%%%%%%%%%%%%%%%%%%%%%%

}  % do not remove

%%% Local Variables: 
%%% mode: latex
%%% TeX-master: "../hadron2011.tex"
%%% End: 

\end{document}